\documentclass[12pt,preprint]{aastex}
\begin{document}

\newcommand{\kms}{\ensuremath{\mathrm{km}\,\mathrm{s}^{-1}}}
\newcommand{\etal}{et al.}
\newcommand{\LCDM}{$\Lambda$CDM}
\newcommand{\ML}{\ensuremath{\Upsilon_{\star}}}
\newcommand{\MLmax}{\ensuremath{\Upsilon_{max}}}
\newcommand{\MLpop}{\ensuremath{\Upsilon_{pop}}}
\newcommand{\MLopt}{\ensuremath{\Upsilon_{acc}}}
\newcommand{\Om}{\ensuremath{\Omega_m}}
\newcommand{\Ob}{\ensuremath{\Omega_b}}
\newcommand{\OL}{\ensuremath{\Omega_{\Lambda}}}
\newcommand{\C}{\ensuremath{\mathfrak{C}}}
\newcommand{\B}{\ensuremath{\mathfrak{B}}}
\newcommand{\Q}{\ensuremath{{\cal Q}}}
\newcommand{\Pop}{\ensuremath{{\cal P}}}
\newcommand{\G}{\ensuremath{{\Gamma}}}
\newcommand{\Lsun}{\ensuremath{L_{\odot}}}
\newcommand{\Msun}{\ensuremath{{\cal M}_{\odot}}}
\newcommand{\mass}{\ensuremath{{\cal M}}}
\newcommand{\Mst}{\ensuremath{{\cal M}_{\star}}}
\newcommand{\Mg}{\ensuremath{{\cal M}_g}}
\newcommand{\Md}{\ensuremath{{\cal M}_d}}
\newcommand{\Mh}{\ensuremath{{\cal M}_h}}
\newcommand{\Sd}{\ensuremath{{\Sigma}_0}}
\newcommand{\fst}{\ensuremath{f_{\star}}}
\newcommand{\Vst}{\ensuremath{V_{\star}}}
\newcommand{\vst}{\ensuremath{v_{\star}}}
\newcommand{\Vg}{\ensuremath{V_{g}}}
\newcommand{\Vb}{\ensuremath{V_{b}}}
\newcommand{\Vh}{\ensuremath{V_h}}
\newcommand{\Vf}{\ensuremath{V_{f}}}
\newcommand{\VNFW}{\ensuremath{V_{200}}}
\newcommand{\shape}{\ensuremath{\mathfrak{T}_{0.6}}}
\newcommand{\norm}{\ensuremath{\sigma_8}}
\newcommand{\gn}{\ensuremath{g_N}}
\newcommand{\csb}{\ensuremath{\mu_0}}
\newcommand{\magsq}{\ensuremath{\mathrm{mag.}\,\mathrm{arcsec}^{-2}}}
\newcommand{\surfdens}{\ensuremath{{\cal M}_{\odot}\,\mathrm{pc}^{-2}}}
\newcommand{\MDAC}{MDAcc}
\newcommand{\HI}{H{\sc i}}


\title{The Rotation Velocity Attributable to Dark Matter at Intermediate
Radii in Disk Galaxies} 

\author{S.S.~McGaugh\altaffilmark{1}, 
W.J.G.~de Blok\altaffilmark{2},
J.M.~Schombert\altaffilmark{3},
R.~Kuzio de Naray\altaffilmark{1},
and J.H.~Kim\altaffilmark{1}} 


\altaffiltext{1}{Department of Astronomy, University of Maryland,
 College Park, MD 20742-2421}
\altaffiltext{2}{Research School of Astronomy \& Astrophysics, Mt. Stromlo 
 Observatory, Australian National University, Cotter Road, Weston Creek, 
 ACT 2611, Australia}
\altaffiltext{3}{Department of Physics, University of Oregon, Eugene,
 OR 97403}

\begin{abstract}
We examine the amplitude of the rotation velocity that can be
attributed to the dark matter halos of disk galaxies, 
focusing on well measured intermediate radii.
The data for 60 galaxies spanning a large range of mass
and Hubble types, taken together, 
are consistent with a dark halo velocity $\log \Vh = \C + \B \log r$ with
$\C = 1.47^{+0.15}_{-0.19}$ and $\B \approx \onehalf$ over the
range $1 < r < 74$ kpc.  The range in $\C$ stems from different 
choices of the stellar mass estimator, from
minimum to maximum disk.  For all plausible choices of 
stellar mass, the implied densities of the dark halos are 
lower than expected from structure formation simulations in \LCDM,
which anticipate $\C > 1.6$.  This problem is not specific to a particular 
type of galaxy or to the innermost region of the halo (cusp or core); the
velocity attributable to dark matter is too low at all radii.
\end{abstract}


\section{Introduction}

The rotation curves of spiral galaxies provide one of the strongest lines of
evidence establishing the need for dark matter in the universe 
(e.g., Rubin, Thonnard \& Ford 1980; Bosma 1981).  An elaborate paradigm for dark
matter has subsequently developed: \LCDM.  More recent work
emphasizes larger scale constraints
(e.g., Tegmark \etal\ 2004; Spergel \etal\ 2006). 
An important question is whether this cosmic paradigm provides a
satisfactory description of spiral galaxies.

A complete explanation of spiral galaxies in \LCDM\ requires a 
comprehensive theory of galaxy formation.  This remains lacking.
Indeed, there are a number of lingering
problems on small (galaxy) scales.  These include the cusp-core problem
(e.g., de Blok \etal\ 2001a), the missing satellite problem (e.g., Moore \etal\ 1999a), 
the dynamical friction time scale problem (Goerdt \etal\ 2006;
Sanchez-Salcedo, Reyes-Iturbide, \& Hernandez 2006), 
and a whole suite of other incongruities and apparent contradictions that arise
on the scale of individual galaxies 
(McGaugh \& de Blok 1998; Sellwood \& Kosowsky 2001).  
In this paper, we will concentrate on the amplitude of the
rotation velocity that is observed for spiral galaxy dark matter halos and that
predicted by \LCDM.

Considerable attention has focused on the density profile inferred 
for the dark matter halos of dwarf and low surface brightness
(LSB) galaxies at small radii (Moore 1994; Flores \& Primack 1994;
C{\^o}t{\'e} \etal\ 2000; Swaters \etal\ 2000; de Blok \etal\ 2001a,b; 
Borriello \& Salucci 2001; Salucci 2001; 
Bolatto \etal\ 2002; Marchesini \etal\ 2002; Salucci \etal\ 2003;
Swaters \etal\ 2003a,b; de Blok \etal\ 2003; Donato \etal\ 2004;
Simon \etal\ 2003, 2005; Gentile \etal\ 2004, 2005; 
Spekkens, Giovanelli, \& Haynes 2005; Kuzio de Naray \etal\ 2006;
Zackrisson \etal\ 2006).
There are good reasons for this:  LSB galaxies are dark matter dominated
down to small radii, making the inference of their dark matter properties
much less sensitive to the stellar mass than in higher surface brightness spirals.
In the most straightforward analyses, the data appear to contradict a basic
prediction of the \LCDM\ structure formation scenario, that dark matter
halos should have cusps at small radii (a density distribution 
$\rho \sim r^{-\gamma}$ with $\gamma \approx 1$:  Dubinski 1994;
Cole \& Lacey 1996; Navarro, Frenk, \& White 1997 (hereafter NFW); 
Tormen, Bouchet, \& White 1997; Moore et al.\
1998, 1999b; Tissera \& Dominguez-Tenreiro 1998; Nusser \& Sheth
1998; Syer \& White 1998; Avila-Reese,
Firmiani, \& Hernandez 1998; Salvador-\'Sole, Solanes,
\& Manrique 1998; Jing 2000; Jing \& Suto 2000; Kull 1999;
Klypin \etal\ 2001; Ricotti 2003; Power \etal\ 2003; Diemand \etal\ 2005).  
The data are apparently more consistent with a nearly constant density core
($\gamma \approx 0$).  Many different solutions
to this problem have been proposed (e.g., El-Zant, Shlosman, \& Hoffman 2001; 
Weinberg \& Katz 2002; Merritt \etal\ 2004; Ma \& Boylan-Kolchin 2004;
Mashchenko, Couchman, \& Wadsley 2006; Tonini, Lapi, \& Salucci 2006).

Interesting constraints also stem from the rotation curve data for high surface 
brightness (HSB) galaxies.  It is frequently possible to obtain quite high
quality data for these bright objects, but the interpretation depends critically on 
the mass ascribed to baryons.  The stellar disks in these objects can be 
massive (e.g., Debatista \& Sellwood 2000; Salucci \& Burkert 2000; 
Weiner, Sellwood, \& Williams 2001; Zavala \etal\ 2003;
Kassin, de Jong, \& Weiner 2006) 
or even maximal (e.g., van Albada \& Sancisi 1986; Kent 1987; 
Palunas \& Williams 2000; Battaglia \etal\ 2006).
Such dynamically dominant disks leave little room for halos with steep inner
density profiles.  For example, Binney \& Evans (2001) find that cuspy halos
are inconsistent with a number of constraints for the Milky Way
(cf.\ Klypin, Zhao, \& Somerville 2002).
On the other hand, it can be argued that the lack of scale length residuals
in the Tully-Fisher relation implies that typically HSB disks are sub-maximal
(e.g., Courteau \& Rix 1999; cf.\ Sellwood 1999).  
If dark matter is dynamically significant in the inner
regions of HSB galaxies, then it may be quite possible for the halos of these
galaxies to have a steep cusp.  The critical issue is the stellar mass.

Constraining the precise inner slope of the dark matter density profile
of an individual galaxy is observationally challenging.  Even if the stellar
mass can be properly treated, a variety of systematic effects might affect the
result (Swaters \etal\ 2003a; de Blok \etal\ 2003; Simon \etal\ 2003, 2005;
Spekkens \etal\ 2005).  There are several distinct problems here: observational
uncertainties, non-circular motions that might cause misestimation of the
potential, and degeneracy between fitting parameters.  

Observational uncertainties of concern here are those that might lead us
to underestimate the steepness of the rise of the rotation curve.
This might result, for example, from beam-smearing of 21 cm observations 
(e.g., van den Bosch \etal\ 2000)
or mis-positioning of the slit during optical observations (e.g., 
Swaters \etal\ 2003a).  Such errors are not a great concern when rotation
curves are observed to be steeply rising, as is typically the case for HSB galaxies
in this sample.  When the observed rotation curve rises more gradually,
there is more potential for concern as to whether this behavior is truly intrinsic
(de Blok, McGaugh, \& van der Hulst 1997; de Blok \& McGaugh 1997).  
This no longer appears to be a significant worry for LSB galaxies, as high 
resolution two dimensional velocity fields (Kuzio de Naray \etal\ 2006) 
largely confirm pervious results 
(e.g., McGaugh, Rubin, \& de Blok 2001; de Blok \& Bosma 2002;
see also Zackrisson \etal\ 2006).   

Non-circular motions can be significant at small radii, in the sense that they
can represent a substantial fraction of the total gravitational potential.
The contributions of circular and noncircular motions sum in quadrature.  
From this, it immediately follows that non-circular motions ($V_r$) 
are only important when comparable to the rotational motion ($V_{\phi}$).  
Simon \etal\ (2005) show 
some cases for which this is the case ($V_r \approx V_{\phi}$), albeit at quite
small radii.  For the sample considered here, which all have two dimensional
velocity fields, $V_r$ is typically small (a few to tens of \kms) compared
to $V_{\phi}$ at the same radius (tens to hundreds of \kms).  We consider only
intermediate and large radii where $V_{\phi}^2 \gg V_r^2$ and the contribution
of non-circular motions to the potential is not significant.  
Galaxies that do not satisfy this criterion are not in the sample.

A more fundamental difficulty is the degeneracy 
between halo fitting parameters, even when the potential is
perfectly well known.  The degeneracy between stellar mass and the core
radius of a pseudo-isothermal halo has long been known (e.g., Kent 1987).
For NFW halos, there is substantial covariance between $c$ and \VNFW\ even
when the stellar mass is fixed (de Blok \etal\ 2001b).  This degeneracy stems 
from the similarity of the inner shapes of NFW halo velocity profiles over
a range of ($c$,\VNFW).  More generally, degeneracy is
unavoidable because of a surplus of parameters.  Mass models require the
specification of a minimum of three parameters per galaxy (one for
the disk and two for the halo) while empirically a single parameter suffices to
describe the data (e.g., Brada \& Milgrom 1999; McGaugh 2004).

In this paper, we adopt a model independent approach.  
We do not fit specific mass models to individual galaxies.
The halo parameters derived from such models
inevitably suffer from parameter degeneracies.  
Rather, we simply ask what range of halo parameters ($c$, \VNFW)
are consistent with the halo velocities implied by observation.
We determine the velocity attributable to the dark matter for 
a broad range of stellar mass estimators. 

We find that the data are consistent with a nearly universal dark matter
halo profile.  There is some scatter about this relation,
and we would not claim that there is in fact a single halo profile
appropriate to all galaxies, but the data are not wildly at odds with 
such a situation.  The slope we find is consistent with previous 
determinations (e.g., Barnes, Sellwood, \& Kosowsky 2004).
If halos have the NFW form, their concentrations and/or masses
(\VNFW) are generally smaller than might nominally be expected in \LCDM.
This is largely consistent with various previous results (e.g., Debatista \&
Sellwood 2000; Weiner \etal\ 2001; Borriello \& Salucci 2001;
Zentner \& Bullock 2002; Alam, Bullock, \& Weinberg 2002; 
McGaugh \etal\ 2003; Bell \etal\ 2003a; Gentile \etal\ 2004; 
Zackrison \etal\ 2006; Portinari \& Sommer-Larsen 2006) 
that suggest that real dark matter halos are less dense than predicted 
by \LCDM.

The data and stellar mass estimators are described in \S 2.
In \S 3 we determine the halo velocity \Vh\ implied by the data for
a range of stellar mass estimates.  The predictions of \LCDM\ models
and their relation to the data are described in \S 4.  The implications of
the result and possible ways to reconcile the apparent discord between
theory and observation are discussed in \S 5.  A summary of our 
conclusions is given in \S 6.  

\section{Data}

\subsection{Sample}

We employ the rotation curve data and mass models compiled by 
Sanders \& McGaugh (2002), as trimmed for quality by McGaugh (2005).  
These data originate from many independent observers, and represent
the accumulated effort of a substantial community of astronomers.
Sample galaxies span a wide range of morphological types, from Sab through
all spiral types to Irr/Im.  Recent work by Noordermeer \etal\ (2005) focussing on 
the earliest (S0/Sa) type disk galaxies appears contiguous with the data used here.
More important than morphological type is the
broad range of physical parameter space that our sample spans.  
It includes galaxies with circular velocities $50 < \Vf < 300\;\kms$,
baryonic masses $3 \times 10^8 < \mass_b < 4 \times 10^{11}\;\Msun$,
scale sizes $0.5 \le R_d \le 13$ kpc, central surface brightnesses 
$19.6 \le \mu_0^B \le 24.2\;\magsq$, and gas fractions $0.07 \le f_g \le 0.95$.
No restriction is made on surface brightness or any other physical parameter.
The sample represents the dynamics of disks in the broadest sense possible
with available data.

All sample galaxies have high quality, extended rotation curves derived
from 21 cm data cubes.  The high precision sample of McGaugh (2005)
consists of 60 galaxies with $\sim 700$ independent, 
resolved rotation velocity measurements, each with formal
accuracy of 5\% or better.  We choose to exclude data with $r < 1$ kpc
where the  controversy over cusp or core occurs (de Blok 2004).
The point here is to compare reliable data with the expectations of \LCDM\ 
models, free from concern over systematic uncertainties, either observational
(such as beam smearing) or physical (e.g., non-circular motions).
These may be serious concerns for constraining the innermost slope of
the density profile, but do not significantly affect the data at intermediate radii. 
Here the data are well resolved, and the observed circular velocity greatly
exceeds the velocity dispersion.  Indeed, strong constraints exist on the
ellipticities of the potentials for some sample galaxies 
(Schoenmakers, Franx, \& de Zeeuw 1997).  Triaxiality of the halo
(e.g., Hyashi \& Navarro 2006) is not relevant to the bulk of
the data here.

\begin{figure}  
\epsscale{1.0}
\plotone{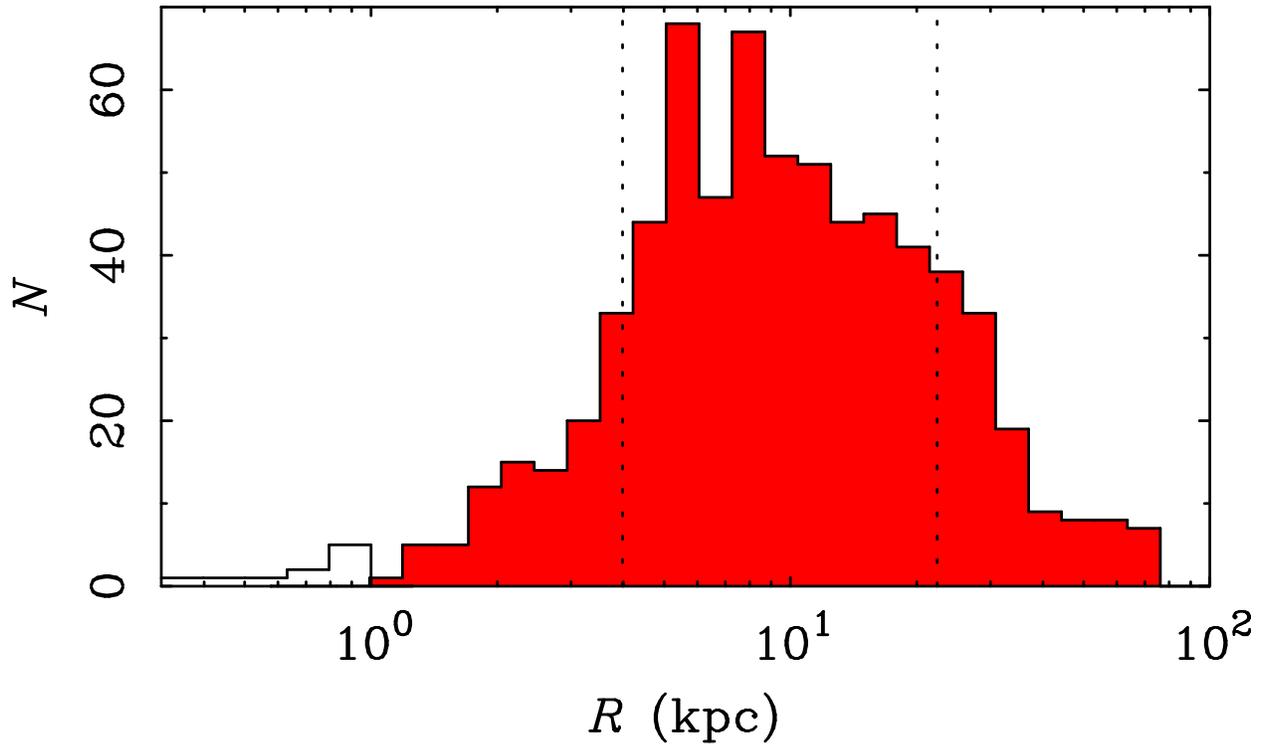}
\caption{The radial distribution of the data from all 60 galaxies.  
Data with $r < 1$ kpc (unshaded portion of histogram) are excluded
from further analysis.  The dotted lines illustrate the
range $4 < r < 22$ kpc containing 68\% of the data, with equal numbers 
on either side of the median (9 kpc).  These data at intermediate radii
dominate the signal and subsequent result.
\label{Rhisto}}
\end{figure}

We exclude the possibility of non-circular motions biasing the results
at small radii by truncating the inner data.
Our result is not sensitive to the choice of the inner truncation radius.
The total radial range sampled is from 1 to 74 kpc, with a median radius
of 9 kpc.  Only a few galaxies have measured
rotation extending to very large radii.  The bulk of the data (68\%) fall in
the range $4 < r < 22$ kpc (Fig.~\ref{Rhisto}).  
Data in this range are not only numerous, but
fall in a range where the systematic concerns discussed above have negligible
impact.  The same result follows if we 
restrict the data to this very conservative range.  

\placefigure{Rhisto}

For many galaxies in the sample, the contribution of the stars to the total
potential can not be ignored.  This dominates all other effects by far.
It is necessary to subtract off the baryonic contribution to reveal the 
velocity due to the halo.  All galaxies have detailed surface 
photometry from which baryonic mass models are derived.  

For the present work, we utilize $B$-band surface photometry, as this is
the only band that all 60 galaxies have in common.  
We are working to place the sample on a uniform $K_s$-band scale, but at
present lack sufficiently deep observations of some of the fainter galaxies.
Using the $K_s$-band data already in hand does not alter the basic result
(see also Verheijen 2001; Verheijen \& Sancisi 2001; Kassin \etal\ 2006).

We make no assumption about the light profile 
(e.g., exponential disk).  The observed light distribution is converted to
a baryonic rotation curve numerically (via GIPSY:  van der Hulst \etal\ 1992).  
This includes the bulge (when present), stellar disk, and \HI\ gas.
The molecular gas distribution is not available in most cases, but this is  
likely to be a small percentage of the baryonic mass in the majority of 
late type galaxies (Young \& Knezek 1989).  
Moreover, the molecular gas generally follows the stellar distribution
(Helfer \etal\ 2003), so is effectively subsumed into the stellar mass-to-light ratio. 
It is the estimation of stellar mass that is most crucial.

\subsection{Stellar Mass}

The observed rotation curves provide a good map of the total gravitational
potential that is due to both baryonic and dark mass.  While the total 
potential is well constrained, dividing it into baryonic and dark mass
components is a degenerate process.  Additional information 
is needed to break this degeneracy.

We explore three methods for estimating the stellar mass-to-light ratio \ML:
maximum disk, \MLmax; stellar population synthesis models, \MLpop; 
and that from the mass discrepancy--acceleration relation, \MLopt.  
This latter in effect treats MOND (Milgrom 1983) fits
(Sanders \& McGaugh 2002) as an empirical phenomenology.
That this phenomenology is present in the data is all that is required,
regardless of the underlying physical reason for it (McGaugh 2004).  

Following McGaugh (2005), we can choose any arbitrary stellar mass 
estimate by scaling each of these prescriptions by a constant 
factor \G, \Pop, or \Q:
\begin{mathletters}
\begin{eqnarray}
\ML = \G \MLmax \\
\ML = \Pop \MLpop \\
\ML = \Q \MLopt.
\end{eqnarray}
\end{mathletters}
This scheme allows us to explore the full range of possible stellar masses
($0 \le \G \le 1$) with three independent estimators for each galaxy.
The population synthesis mass-to-light ratio \MLpop\ is estimated from 
the observed $B-V$ color (as tabulated by McGaugh 2005)
and the models of Bell \etal\ (2003b):
\begin{equation}
\log \MLpop = 1.737 (B-V) -0.942
\end{equation}
Very similar results follow from other models (e.g., Portinari \etal\ 2004).
The chief uncertainty of this method is the IMF.  $\Pop = 1$ corresponds
to the nominal ``scaled Salpeter'' IMF adopted by Bell \etal\ (2003b), which
for our purposes is nearly indistinguishable from the Kroupa (2002) 
IMF adopted by Portinari \etal\ (2004; see also Portinari \& 
Sommer-Larsen 2006).  To consider the effects of a heavier
or lighter IMF, we simply scale by \Pop\ (see discussion in
Bell \& de Jong 2001).

\section{Empirical Halo Estimates}

We estimate the circular velocity provided by dark matter
by subtracting the baryonic component from the observed rotation curves 
for each choice of \ML:
\begin{equation}
\Vh^2(r) = V^2(r) - \Vb^2(r).
\end{equation}
Here $\Vh(r)$ is the rotation velocity attributed to the dark matter, 
$V(r)$ is the observed rotation velocity, and $\Vb(r)$ is the rotation
velocity attributed to baryons.  This depends on \ML\ through
$\Vb^2 = \ML \vst^2 + \Vg^2$, where \vst\ is the rotation velocity
for $\ML = 1\;\Msun/\Lsun$.  The \HI\ gas component, 
corrected to account for helium and metals, gives \Vg.  Molecular gas generally
traces the stellar distribution (Helfer \etal\ 2003) and is subsumed into \ML.

\begin{figure}  
\epsscale{1.0}
\plotone{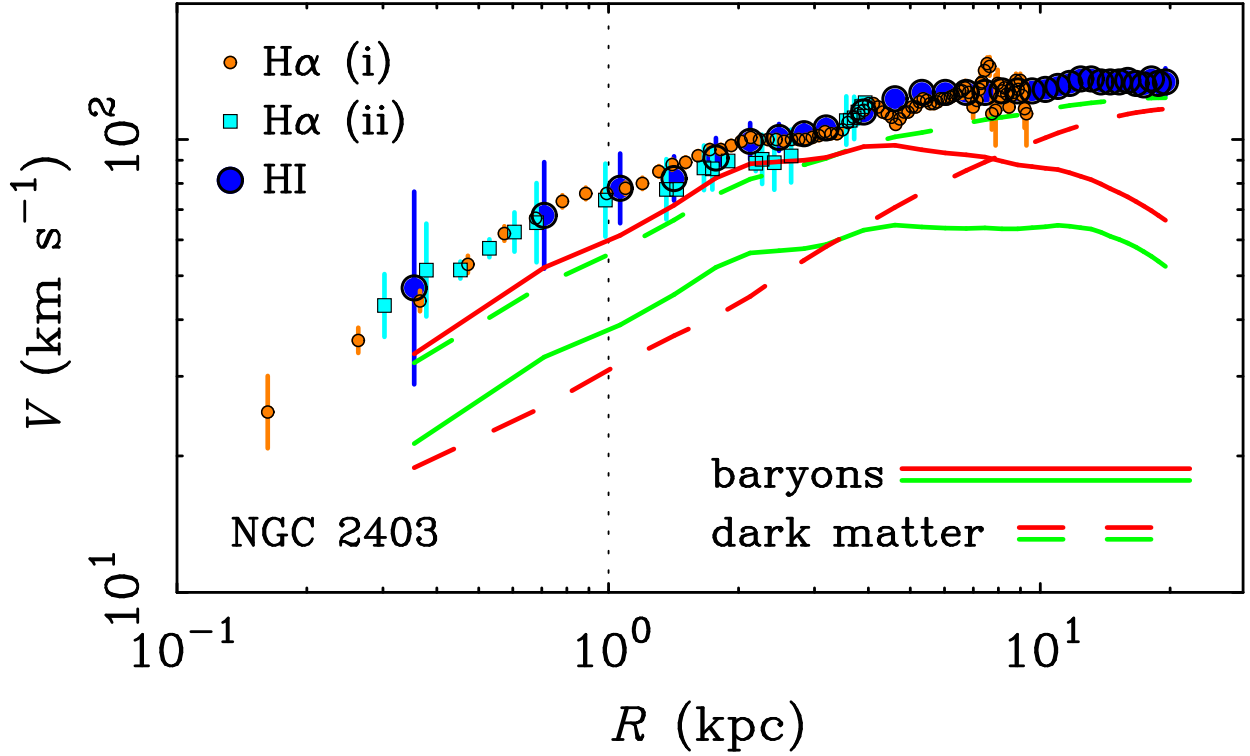}
\caption{The rotation curve and mass model of NGC 2403 in log-log space.
The \HI\ data are from Begeman (1987; see also Fraternali \etal\ 2002), 
H$\alpha$ (i) from Blais-Ouellette \etal\ 
(2004), and H$\alpha$ (ii) from Daigle \etal\ (2006).  The H$\alpha$ data are shown
to illustrate the degree to which independent data sets agree, but only the HI data
are used in the analysis.  The solid lines show the 
mass model \Vb\ of the baryons for two choices of \ML, $\G = 0.9$ (upper) and
$\Pop = 1 $ (lower).  The dashed lines show the resulting halo rotation 
curves \Vh\ for each \ML.  The upper dashed line corresponds to the lower 
baryonic line and vice-versa.  The dotted line at $r = 1$ kpc demarcates
where the innermost data are ignored.
\label{N2403log}}
\end{figure}

\placefigure{N2403log}

An example rotation curve is shown in Fig.~\ref{N2403log}.  
As we increase (decrease) \ML\ by varying \G, \Pop, and \Q, 
the stellar contribution to the total rotation increases (decreases)
and the corresponding contribution from the dark matter 
decreases (increases).  Though the details of each realization depend
on the \ML\ estimator and the relative importance of gas (which does
not vary), it is a tolerable approximation to imagine the
dark and baryonic lines in Fig.~\ref{N2403log} 
sliding up and down see-saw fashion.

\placefigure{VhQboth}

In order to restrict our attention to the best data, we have excised individual
points with uncertainty in excess of 5\%.  In many cases, the formal precision
is rather greater than this limit.  While this leaves plenty of data to analyze, 
various astrophysical considerations make it unrealistic to consider those data
with small error bars as accurate estimates of the circular velocity of the potential 
(see discussions in McGaugh \etal\ 2001; de Blok \etal\ 2003).  We therefore
give uniform weight to the data which pass the 5\% criterion.

A more serious uncertainty than that in the data is that in \ML.  
The estimate of stellar mass is the single dominant source of uncertainty in
this analysis.  The systematic nature of this uncertainty makes it impossible
to propagate it in a meaningful way.  Hence we simply examine the results of
adopting each \ML\ estimator, leaving it to the reader's judgement which case
might seem best.  Fortunately, the basic result is fairly insensitive to this choice.

In Fig.~\ref{VhQboth} we show $\Vh(r)$ for 
all 60 galaxies plotted together for the case
of $\Q = 1$.  Remarkably, the inferred dark matter halos of all these galaxies
fall more or less on top of each other.  This is not to say that the dark matter
halos of all galaxies are the same.  Quite the contrary, it is possible to
observationally distinguish between the dark matter halos of individual
galaxies (Sancisi 2004; McGaugh 2004; Barnes \etal\ 2004; Simon \etal\ 2005).  
Nonetheless, when plotted together in this fashion, the apparent 
variation is remarkably small.

Taken together, the data suggest a relation
\begin{equation}
\log \Vh = \C + \B \log r.  \label{singlehalo}
\end{equation}
Fitting the data with $\Q = 1$ gives $\C = 1.48 \pm 0.01$ and 
$\B = 0.49 \pm 0.01$.
While we would not claim that this is a universal halo that all galaxies
share, it does give a tolerable approximation of the velocity
attributable to dark matter.

We can extend this approach to other mass-to-light values by varying 
\G, \Pop, and \Q.  The cases $\G = 1$, $\Pop = 1$, $\Q = 1$ and $\G = 0.4$
(the latter as suggested\footnote{These authors phrase the disk contribution 
in terms of the velocity at 2.2 scale lengths.  Since $\mass \propto V^2$,
$V_{2.2}/V_{tot} = 0.63$ corresponds to $\G = 0.4$.}
by Bottema 1993 and Courteau \& Rix 1999) are
illustrated in Fig.~\ref{DMscale}.  This figure is like 
Fig.~\ref{VhQboth} for the various mass estimators, but with shading to
illustrate the  density of data.  The darker the shading, the more sharply
defined the relation between \Vh\ and radius.

There is some covariance between the slope \B\ and
intercept \C.  Our purpose here is to constrain the
amplitude of the velocity that can be attributed to dark matter.  In order
not to obscure the basic result with uncertainties introduced by covariance,
we fix the slope to $\B = 1/2$, consistent with our data
(see Figs.~\ref{DMhistoG1} -- \ref{DMhistoG04})
and those of Barnes \etal\ (2004).
Moreover, the slope does not vary appreciably when left free; the important
information is contained in the amplitude.

\placefigure{DMhistoG1}
\placefigure{DMhistoP1}
\placefigure{DMhistoQ1}
\placefigure{DMhistoG04}

Fixing $\B = 1/2$ changes the $\Q = 1$ best fit intercept to $\C = 1.47$.
This number encodes, in the simplest possible way, the amplitude of the dark 
matter velocity.  Fig.~\ref{scatterplot} show how this amplitude varies as 
we change \ML.  For $\Pop = 1$, the amplitude barely changes: $\C = 1.49$.  
For the full range of possibilities, from minimum to maximum disk, we find
$\C(\G = 0) = 1.62$ and $\C(\G = 1) = 1.28$.  In other words, the 
amplitude of the velocity provided by the dark matter falls in the
range $\C = 1.47^{+0.15}_{-0.19}$ for $\B = 1/2$.

\placefigure{scatterplot}

One can also see in Fig.~\ref{scatterplot} variation in the scatter about the mean.  
Remarkably, the scatter is minimized for $\Q = 1$, the MOND 
mass-to-light ratios.  This is also the choice of \ML\ which minimizes the
scatter in the baryonic Tully-Fisher relation (McGaugh 2005) and, by construction,
that in the mass discrepancy--acceleration relation (McGaugh 2004).  Irrespective
of the physics that causes it, this implies that MOND is an important 
empirical organizing relation that is relevant even to dark matter.
The scatter is larger for other mass estimators, but the basic result
encapsulated by \C\ is the same.

\section{\LCDM\ Halos}
 
Cosmological simulations lead to mass distributions for dark matter
halos that have gravitational potentials which can be analytically approximated as
\begin{equation}
\Vh^2(r) = \VNFW^2 \left[\frac{\ln(1+cx) - cx/(1+cx)}
{x[\ln(1+c)-c/(1+c)]}\right],  \label{NFWformula}
\end{equation}
where $c = R_{200}/R_s$ and $x=r/R_{200}$ (NFW).  The radius 
$R_{200}$ encloses a density 200 times the critical density
of the universe; $V_{200}$ is the circular velocity of the potential 
at $R_{200}$.   

The radius $R_{200}$ is, crudely speaking, the virial radius in a critical
density universe.  For current \LCDM\ parameters, the virial radius is more
nearly at $R_{100}$ (e.g., Eke, Navarro, \& Steinmetz 2001).  More recently,
Navarro \etal\ (2004) have suggested that a density profile with a continuously
running slope provides a better fit to simulated halos than the original NFW formula.
From an observational perspective, these forms are virtually indistinguishable.  
We therefore retain the NFW formulation for its familiarity and its
well quantified relation to cosmological parameters 
(NFW; McGaugh, Barker, \& de Blok 2003).

One interesting aspect of the data is that there is no clear tendency for the dark
halo rotation curves to flatten out.  The combination of disk plus halo is often
close to flat, and as the disk contribution is declining, it is natural to associate
the flat rotation velocity with the halo.  This is, however, an oversimplification.  
In most cases, the inferred dark matter contribution to the velocity is still rising
at the last measured point, and it is not clear that the observed flat velocity is
truly representative of the halo.  

In the context of simulated halos, the lack of a flat halo rotation
velocity may help the models.  NFW halos do not possess a significant range
of radii where $\rho \propto r^{-2}$, as needed to produce flat rotation curves.
Presumably, the observed range suffices only to cover the rising portion of the
halo, with a suggestion of a turn-over in only a few sufficiently extended cases.
It remains an open question why the disk and halo conspire to produce
nearly flat rotation curves over the range observed.

It is suggestive that the fitted slope $\B \approx 1/2$ is that expected for the 
inner portion of NFW halos.  It is tempting to conclude that halos are indeed 
NFW, and we will use this to compare the observed and predicted amplitudes \C.
An important caveat is that it is not straightforward to match the radial scale
of simulated halos to that of real galaxies.  It is for this reason that the 
amplitude \C\ provides the most robust comparison:  it uses all the data
to set the velocity scale, in effect providing an integrated measure that is
not sensitive to systematic effects in the way the innermost slope may be.

\begin{figure}
\epsscale{0.7}
\plotone{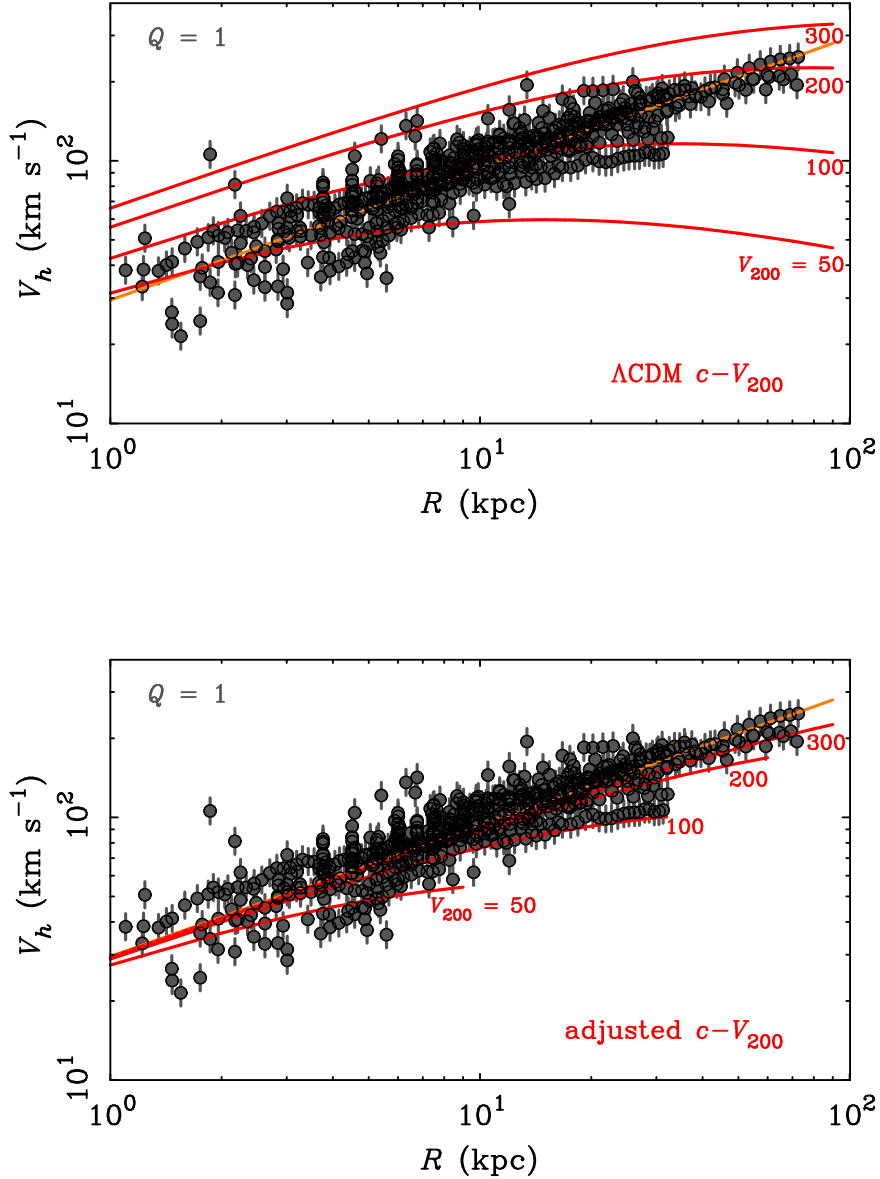}
\caption{The rotation velocity \Vh\ attributable to the dark matter halos of 60
galaxies (points) for $\Q = 1$ stellar masses.  Taken together, these data
form a narrow sequence that is well described by
$\log \Vh = 1.47 + 0.5 \log R$ (straight line).
Also shown are galaxy mass NFW halos with $\VNFW = 50$, 100, 200, and 
300 \kms\ (curved lines).  In the top panel, these are drawn for the 
concentrations predicted for the vanilla \LCDM\ parameters of 
Tegmark \etal\ (2004).  In the lower panel, the $c-\VNFW$ relation 
has been adjusted to match the data.
\label{VhQboth}}
\end{figure}

The slope of the data is, if anything, too close to a constant $\B = 1/2$.  
NFW profiles have this slope at small radii, though the exact slope the
simulations predict is still debated (e.g., Ricotti 2003; Power \etal\ 2003; 
Navarro \etal\ 2004; Diemand \etal\ 2005).  
Irrespective of this detail, all simulations agree
that the inner slope gradually transitions to a steeper fall off (becoming
$r^{-3}$ at large radii).  The consequence is that there should be some 
curvature apparent in $\Vh(r)$ over the radial range sampled by the data.  
This is not obvious in the observations (Fig.~\ref{VhQboth}).
This difference between simulations and data is sufficiently large that the 
difference between the various proposed analytic formulas is a rather small concern.
The use of individually simulated halos (e.g., Hayashi \etal\ 2004; cf.\ 
de Blok 2005) in preference to the analytic approximations should not make a 
difference to the radii of interest here, where the deviations between the various
realizations of the simulation results are small.

In addition to the data, 
the upper panel of Fig.~\ref{VhQboth} shows the predicted shape
and amplitude of NFW rotation curves for a range of \VNFW.  
Presuming galaxy halos span a range of mass comparable to that suggested
by their luminosities, one would not only expect more curvature than 
observed, but also a wider spread in \Vh\ (compare, for example, the halos
with $\VNFW = 50$ and $200\; \kms$).  That the data concentrate in this
plane more than anticipated might suggest that we may err in presuming that halo
masses span the same range of mass as the observed luminosity, or that the
simulated $c$-\VNFW\ relation built into the lines in the upper panel of
Fig.~\ref{VhQboth} may not be precisely correct.

One must bear in mind, of course, that the data only extend over a finite range
of radii, and do not generally show the inevitable maximum and decline in \Vh.  
Though it is difficult to assign a specific halo to a particular galaxy, taken together
the data are suggestive of an envelope composed of halos of a range of masses.
If we had complete knowledge of the potentials out to arbitrary radii, we might
see individual halos peeling away from this envelope as illustrated by the lines
in the bottom panel of Fig.~\ref{VhQboth}. 

For comparison with the data, we can ask what we would expect to find from 
a \LCDM\ model observed in the same way as the data.  We sample a series of 
halos with $50 \le \VNFW \le 300$ 
(e.g., the lines in the top panel of Fig.~\ref{VhQboth}).  Model
data points are allowed to fill the space available to them within the observed
range of radii.  The model points are uniformly sampled and weighted since
we have no knowledge of the underlying halo mass distribution, other than
that it gives rise to the broad observed range of luminosities and rotation velocities.  
Since no turn-over ($V_{max}$ of the halo)
is observed in the data, the maximum radius
to which halos are sampled is truncated along the lower edge of the observed
envelope, as defined by a line with slope 1/2 and intercept 1.33.
The result of fitting the model halos is $\C = 1.66$ and $\B = 0.48$ 
for the vanilla \LCDM\ parameters of Tegmark \etal\ (2004).  
For the third year WMAP parameters (Spergel \etal\ 2006), \C\ 
decreases to 1.61.  This normalization is too large, being consistent with 
the observed value only in the limit of zero stellar mass ($\C = 1.62$).

The $c$-\VNFW\ relation can be adjusted in order to match the NFW model 
to the observed amplitude of the data.  This is illustrated in the lower
panel of Fig.~\ref{VhQboth} by lines representing NFW halos with parameters
chosen to match the observed normalization \C.  In effect, this provides an
empirical estimate of the $c$-\VNFW\ relation.
We can do this for the value of \C\ implied by any choice of \ML.  
To get the zero point amplitude (\C) for NFW,
let us rewrite equation (\ref{NFWformula}) in log-log space.
In the limit\footnote{The limit $x \rightarrow 0$ might seem like a dubious 
approximation since we have excluded the data at $r < 1$ kpc.  However, it
is in fact quite a good approximation both theoretically (since $x = r/R_{200}$
and we are concerned with data at a small fraction of the virial radius) and
empirically (since the observed slope is close to limiting case of 1/2).}
$x \rightarrow 0$ we have: 
\begin{equation}
\log \Vh = \onehalf \left[\log\VNFW+\log[g(c)]+\log(h/2)\right] +\onehalf \log r,
\end{equation}
where
\begin{equation}
g(c) = \frac{c^2}{\ln(1+c)-c/(1+c)}
\end{equation}
and the factor $h = H_0/100\;\textrm{km}\;\textrm{s}^{-1}\;\textrm{Mpc}^{-1}$ 
appears because $\VNFW = h R_{200}$
when \VNFW\ is in \kms\ and $R_{200}$ is in kpc (NFW).  Written this way,
we recognize $\log \Vh = \C + \B \log r$ with $\B = 1/2$.  We can thus relate
NFW parameters to the data through the observed \C:
\begin{equation}
\log \VNFW = 2 \C - \log[g(c)] - \log(h/2)
\end{equation}
(the data assume $h = 0.75$).
In this fashion, we can generate the range of
($c$, \VNFW) that are acceptable to the data (Fig.~\ref{NFWcV}).  
Note that we do not specify
the particular values of these parameters for any given galaxy.  Rather, we
constrain the range of $c$-\VNFW\ parameter space from which the observed
halos may be drawn.

The parameters of NFW halos in simulations are not arbitrary.
There is a correlation between $c$ and \VNFW\ (NFW) with fairly 
modest\footnote{Note that the scatter found by Col{\'{\i}}n \etal\ (2004)
is half that found by Bullock \etal\ (2001).  We adopt the more conservative
(larger) scatter here.  Adopting the smaller scatter would simply make
the predicted bands in Fig.~\ref{NFWcV} $2 \sigma$ wide.}
scatter (Jing 2000; Bullock \etal\ 2001; Col{\'i}n \etal\ 2004).
Moreover, the typical concentration depends on the cosmology (NFW).
For a $10^{12}\;\Msun$ halo, McGaugh \etal\ (2003) found that the mean
\begin{equation}
c = 1.88+23.9 \sigma_8 \shape  \label{concosmo}
\end{equation}
where
\begin{equation}
\shape = \Om^{0.6} h e^{(\Ob - \sqrt{2h} \frac{\Ob}{\Om})} -0.32(n_s^{-1}-1).
\end{equation}
NFW found that the density of halos depends on the density of the universe 
at the time of collapse.  This depends on \Om\ and the amplitude of the 
power spectrum at the appropriate scale, as encapsulated by the above formula.  
The concentration varies slowly with scale \VNFW\ (NFW).

For illustration, we adopt \LCDM\ parameters from the Tegmark \etal\ (2004)
and Spergel \etal\ (2006).  We use the 6-parameter `vanilla' fit of 
Tegmark \etal\ (2004), and the results of the 3 year WMAP-only fit from
Spergel \etal\ (2006).  While superficially these may appear similar, the
concentrations of dark matter halos are quite sensitive to cosmic parameters,
so there is a noticeable difference.  

\placefigure{NFWcV}

The observed and predicted $c$-\VNFW\ bands are illustrated in Fig.~\ref{NFWcV}.
For the scatter in \LCDM\ concentrations, we adopt $\sigma_{\log c} = 0.14$ 
(Bullock \etal\ 2001).  Similarly, we illustrate the $\pm 1\,\sigma$ range
of observed halos (i.e., the band that contains 68\% of the halos from the
histograms in Figs.~\ref{DMhistoP1} and \ref{DMhistoQ1}).  
The predicted and observed bands do not match up well,
having only a modest range of overlap at surprisingly small \VNFW.

\section{Discussion}

The cusp-core problem in dwarf and LSB galaxies 
is well known.  Inspection of Fig.~\ref{NFWcV} reveals a more general problem
with the densities of the dark matter halos of all spiral galaxies.
The data are consistent with a swath of $c$-\VNFW\ space that does
not parallel the predictions of \LCDM.  The acceptable regions of halo
parameter space diverge as mass increases.

This result can not be attributed to inadequate resolution or slit positioning
errors.  All of the data are derived from resolved two dimensional velocity fields.
We have intentionally excluded data from the innermost region where
the resolution might be questioned.  Significant non-circular motions could
be seen directly in the velocity fields.
The result does not change if we exclude\footnote{This result
is slightly less restrictive than the maximum halo acceleration limit
(Brada \& Milgrom 1999; McGaugh 2004) because we ignore here $r = 0$
where NFW halos provide their maximum acceleration.} 
even more of the inner data (see Fig.~\ref{Rhisto}; 
imagine truncating Fig.~\ref{VhQboth} at 2 or 3 kpc).
It is the velocity attributable to dark matter at well resolved intermediate
radii that is lower than expected in \LCDM.

Let us examine the implications of this result, and possible ways out, starting
with the theoretical expectations.  Two fiducial \LCDM\ models are illustrated
in Fig.~\ref{NFWcV}.  The model of Tegmark \etal\ (2004) is a good 
representation of standard vanilla \LCDM, predicting very nearly the same 
swath of $c$-\VNFW\ parameter space as other plausible choices for 
cosmological parameters, going back to the early work of NFW.
The WMAP three year results (Spergel \etal\ 2006) noticeably lower the predicted
halo concentrations.  This is due to slight changes in all of the relevant parameters. 
The dominant effect is from the lower power spectrum normalization 
$\norm = 0.90 \rightarrow 0.74$, a result anticipated both by cluster (e.g.,
Rosati, Borgani, \& Norman, 2002; Diego \etal\ 2003) 
and galaxy work (Zentner \& Bullock 2002;
McGaugh \etal\ 2003; van den Bosch, Mo, \& Yang 2003).
The tilt ($n = 0.99 \rightarrow 0.95$) and the lower matter density
($\Om = 0.27 \rightarrow 0.24$) also contribute.  
As we add in large scale constraints beside WMAP, the parameters
tend\footnote{See http://lambda.gsfc.nasa.gov/product/map/current/parameters.cfm}
to drift back towards those of Tegmark \etal\ (2004).

The lower concentrations implied by the more recent WMAP results do help.
Ignoring the trend with \VNFW\ for the moment, equation (\ref{concosmo}) 
implies $c = 7.6$ for the cosmic parameters of Tegmark \etal\ (2004), whereas 
the new WMAP data (Spergel \etal\ 2006) imply $c = 5.9$.  This difference 
is quite pronounced in rotation curves.  The revised
WMAP results at least bring cosmic data into the right ballpark:  
McGaugh \etal\ (2003) found $c \approx 5.7$ for dark matter dominated galaxies,
though it should be noted that this typical value ignores problematic 
cases with $c < 2$ or for which NFW simply does not fit.

The problem here is not so much the mean concentration, but the run of 
concentration with halo mass.  Concentrations are acceptable for low mass
halos (those with $\VNFW \lesssim 100\kms$).  They become increasingly 
discrepant with increasing mass.  By the time we reach
$\VNFW = 163\;\kms$ ($10^{12}\;\Msun$), the data imply $c = 3$.  
This is nominally the mass scale of the Milky Way (e.g., Wilkinson \& Evans
1999; Sakamoto, Chiba, \& Beers 2003), 
and the low inferred concentration is consistent with both Milky Way constraints
(Binney \& Evans 2001; Flynn \etal\ 2006) and those for comparable galaxies
(e.g, Weiner \etal\ 2001).  To obtain such a
low concentration from initial conditions would require absurd changes to
cosmology: $\norm \approx 0.3$ or $\Om \approx 0.05$.

This basic result is virtually independent of the choice of mass-to-light ratio
estimator.  The distribution of \Vh\ is not as sharply peaked for $\Pop = 1$ as
it is for $\Q = 1$, but the mean is nearly the same, so the same result is
obtained.  Worse, changing \Pop\ (or \Q\ or \G) simply shifts the bands left
(for $\Pop >1$) or right (for $\Pop < 1$) in Fig.~\ref{NFWcV}.  
It does little to alter the
differing slope of the observed and predicted $c$-\VNFW\ bands.  

In order to fix the mismatched slopes with \ML, we would have to make the
stellar mass indicator a systematic function of \VNFW.  Specifically, \ML\ would
have to get progressively smaller as halo mass increases.  This is
quite contrived.  There is no evidence to suggest that the stellar populations 
of massive spirals are systematically lighter than those of dwarfs.  Indeed,
such a contention can likely already be excluded by the need for a substantial
stellar contribution to the mass in at least some high surface brightness galaxies
(e.g., Debattista \& Sellwood 2000; Weiner \etal\ 2001; Kassin \etal\ 2006;
Battaglia \etal\ 2006).  
In contradiction to this, Jimenez, Verde, \& Oh (2003) argue that rotation 
curve fits tolerate the expected concentrations for NFW halos.  However, this
comes at the expense of unrealistic \ML.  The $I$-band \ML\ distribution of
Jimenez \etal\ (2003) is bimodal (their Fig.~7), with roughly half of the galaxies
centered on $\ML \approx 0.2\Msun/\Lsun$.  So they in fact find the same
result, that mass-to-light ratios need to
be quite low in order to be reconciled with the expected NFW concentrations.
A velocity-variable IMF might achieve this, but it would introduce an 
unreasonable amount of scatter into the baryonic Tully-Fisher relation 
(McGaugh 2005).  

\begin{figure}  
\epsscale{1.0}
\plotone{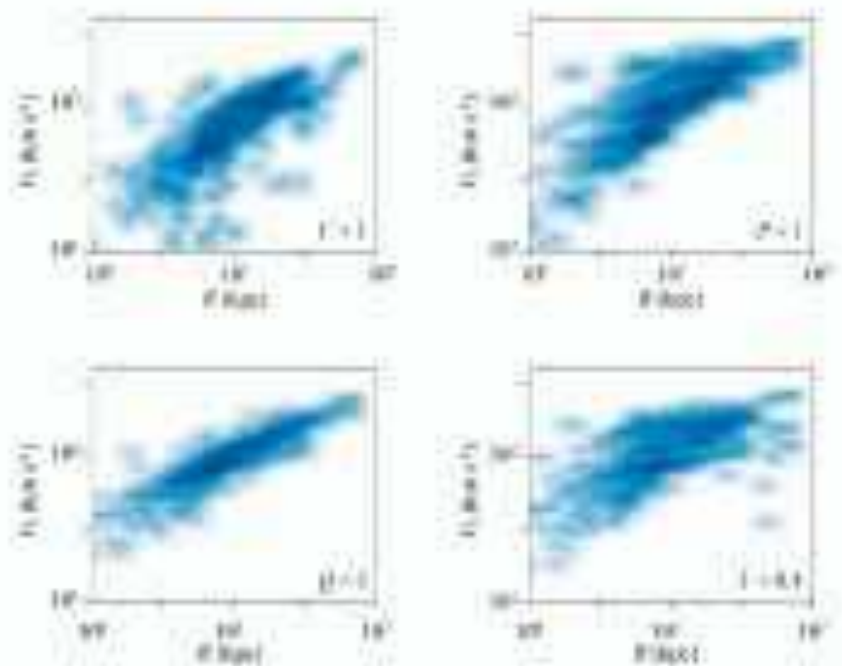}
\caption{Dark halo velocities, as per Fig.~\ref{VhQboth}.  
The data are represented here
as a grayscale to show the density of the data.  Darker shades
representing regions of higher density; lighter shades those of
low density.  In regions of very low data density, individual points
are shown as dots.  While the eye may be attracted to
outliers in Fig.~\ref{VhQboth}, here it is clear how sharply 
defined $\Vh(r)$ is.  Several choices of mass-to-light ratio are illustrated:
maximum disk ($\G =1$, top left); stellar population synthesis ($\Pop = 1$,
top right); MOND ($\Q = 1$, bottom left -- the same as in Fig.~\ref{VhQboth}); 
and sub-maximal (Bottema 1993) disks  ($\G = 0.4$, bottom right).  
\label{DMscale}}
\end{figure}

\begin{figure}  
\epsscale{1.0}
\plotone{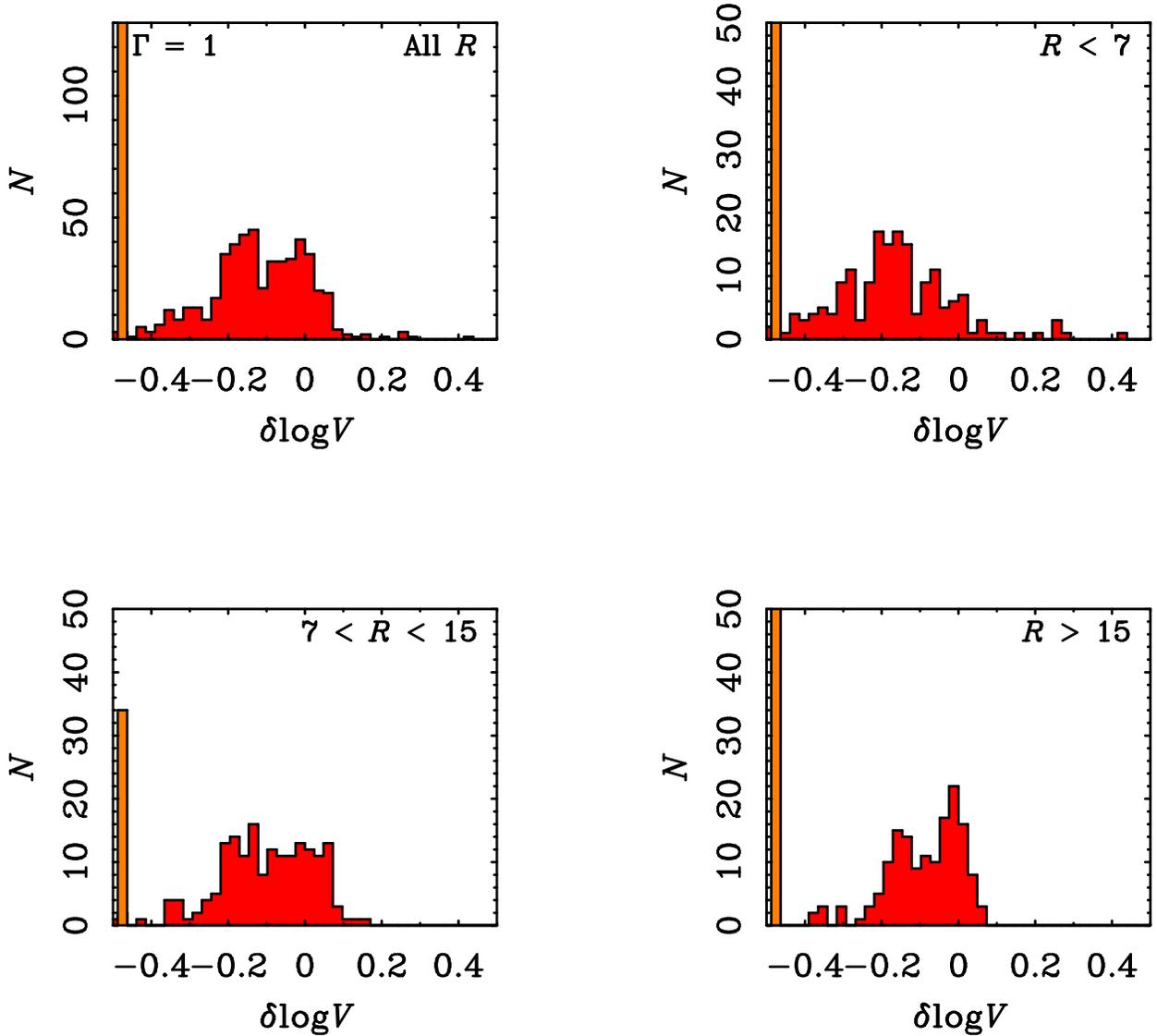}
\caption{The distribution about the line fit to the $\Q = 1$ case
for maximum disk mass-to-light ratios ($\G = 1$).  
In essence, this is the cross-section of the corresponding 
pane in Fig.~\ref{DMscale}.  All the data
are shown in the first panel, with subsequent panels showing the
data falling in radial bins containing approximately equal numbers
of data points.  Scatter causes some individual points
to imply a negative dark mass.  These are placed in the bin at
the left edge of the plot.  The variation of the mean $\delta \log V$
illustrates how the characteristic amplitude of the dark matter velocity
\C\ shifts with \ML, becoming lower in this case.  The radially binned plots
illustrate how the distribution varies.  That the centroid in these bins
does not vary much shows the approximate constancy of the slope \B.
\label{DMhistoG1}}
\end{figure}

\begin{figure}  
\epsscale{1.0}
\plotone{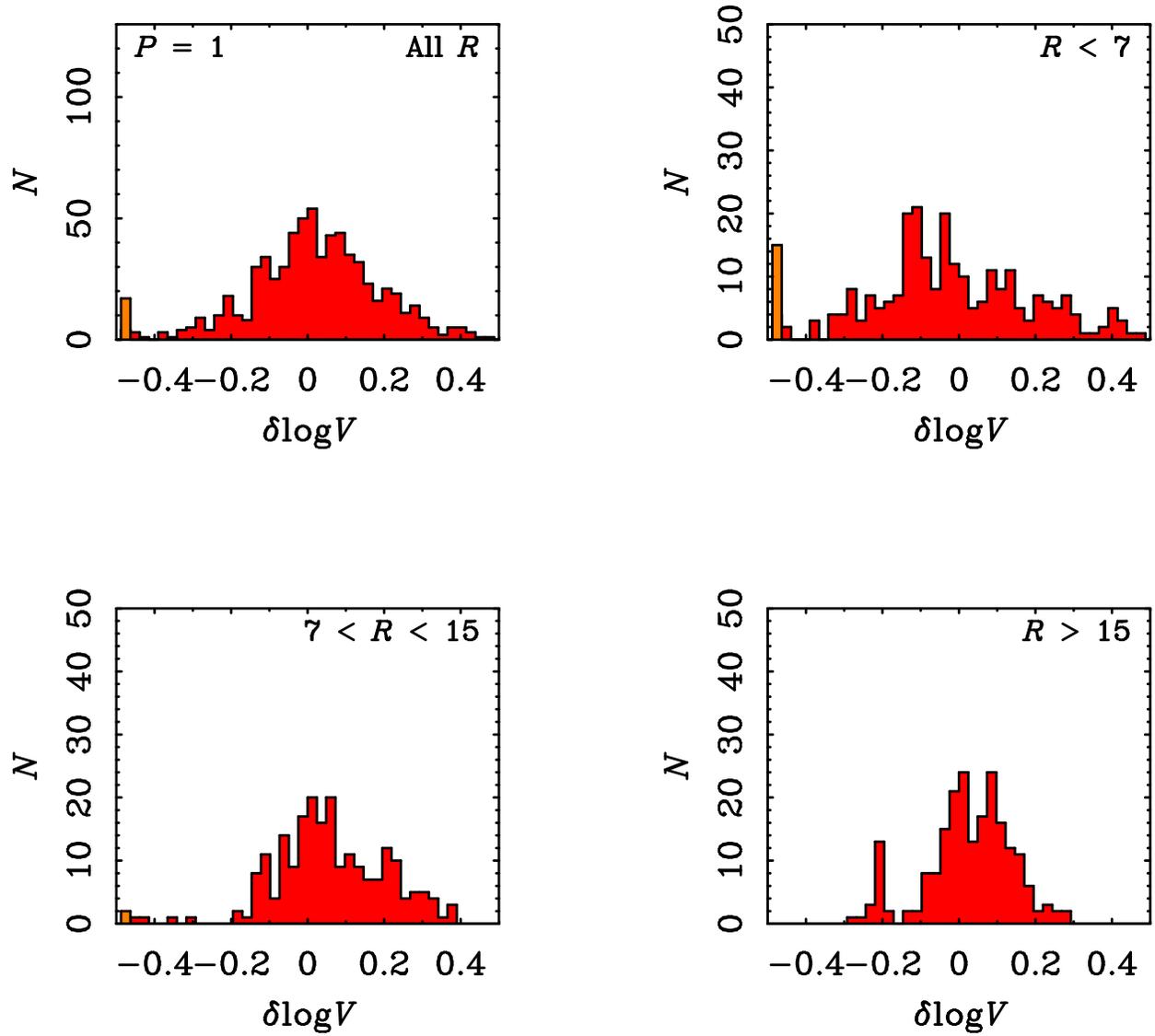}
\caption{As per Fig.~\ref{DMhistoG1}, 
but for population synthesis stellar mass estimates ($P=1$).
\label{DMhistoP1}}
\end{figure}

\begin{figure}  
\epsscale{1.0}
\plotone{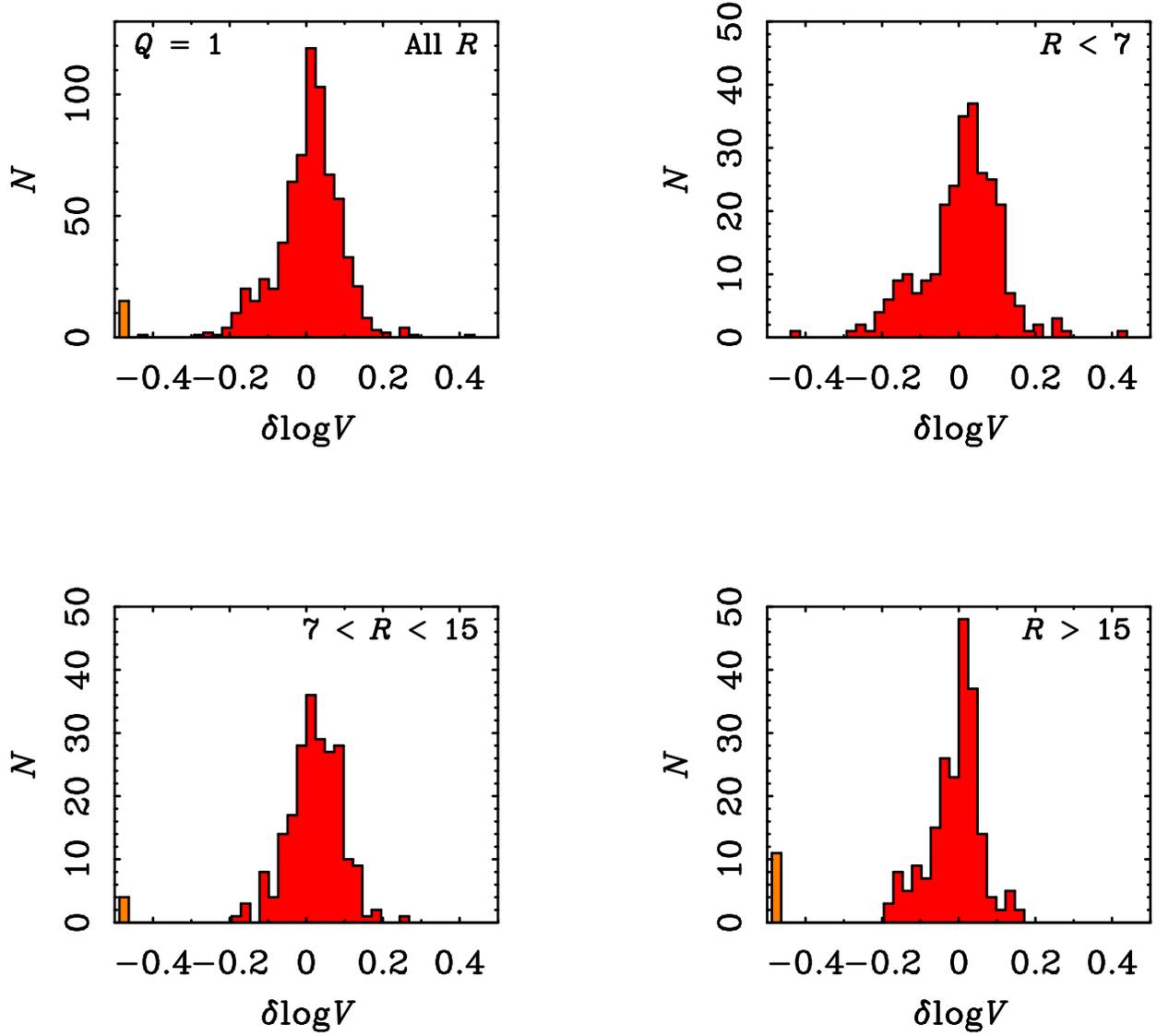}
\caption{As per Fig.~\ref{DMhistoG1}, 
but for MOND ($Q=1$) mass-to-light ratios.
Note how sharply defined the peak in \Vh\ is for this choice of \ML.
\label{DMhistoQ1}}
\end{figure}

\begin{figure}  
\epsscale{1.0}
\plotone{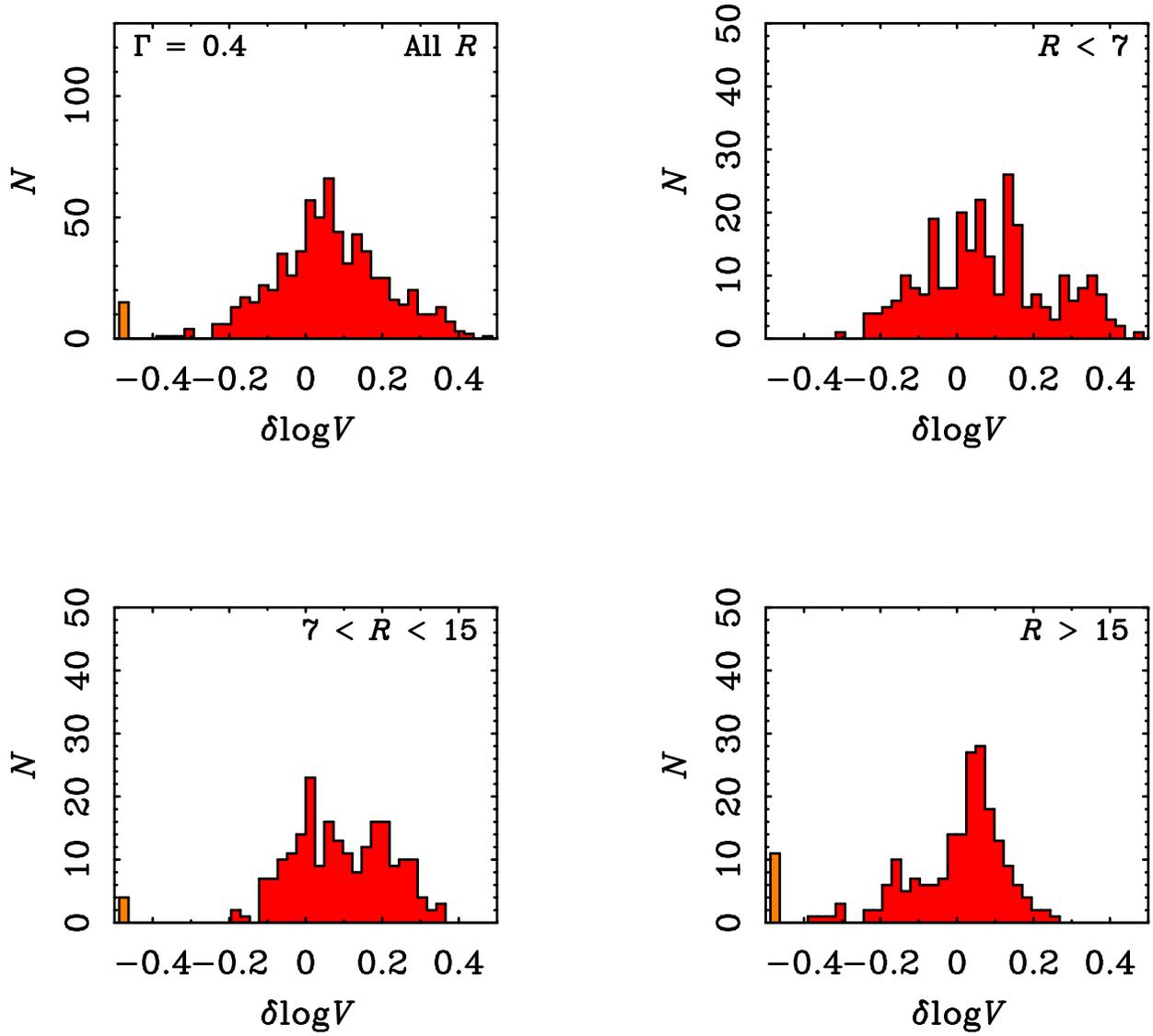}
\caption{As per Fig.~\ref{DMhistoG1}, 
but for sub-maximal disks with $\G = 0.4$.
\label{DMhistoG04}}
\end{figure}

\begin{figure}
\epsscale{1.0}
\plotone{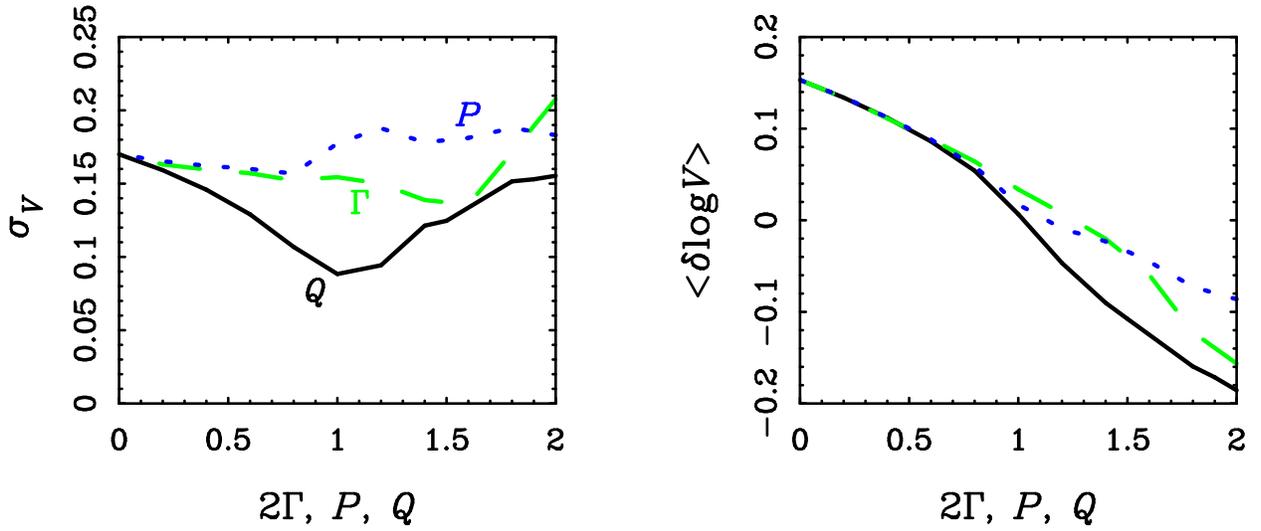}
\caption{Variation of the scatter (left) and mean zero point velocity 
($\Delta \C$ from $\Q = 1$; right) with stellar mass prescription scaling
factor (\Q: solid line; \Pop: dotted line; \G: dashed line).
A clear minimum in the scatter occurs at $\Q = 1$.  The variation for
$\Q > 1$ is not smooth as individual galaxies are excluded as the 
prescription causes them to exceed maximum disk.  Note that for low
stellar masses (\Q, $\Pop < 0.8$, $\G < 0.4$) the different prescriptions
give indistinguishable mean \C.
\label{scatterplot}}
\end{figure}

\begin{figure}
\epsscale{1.0}
\plotone{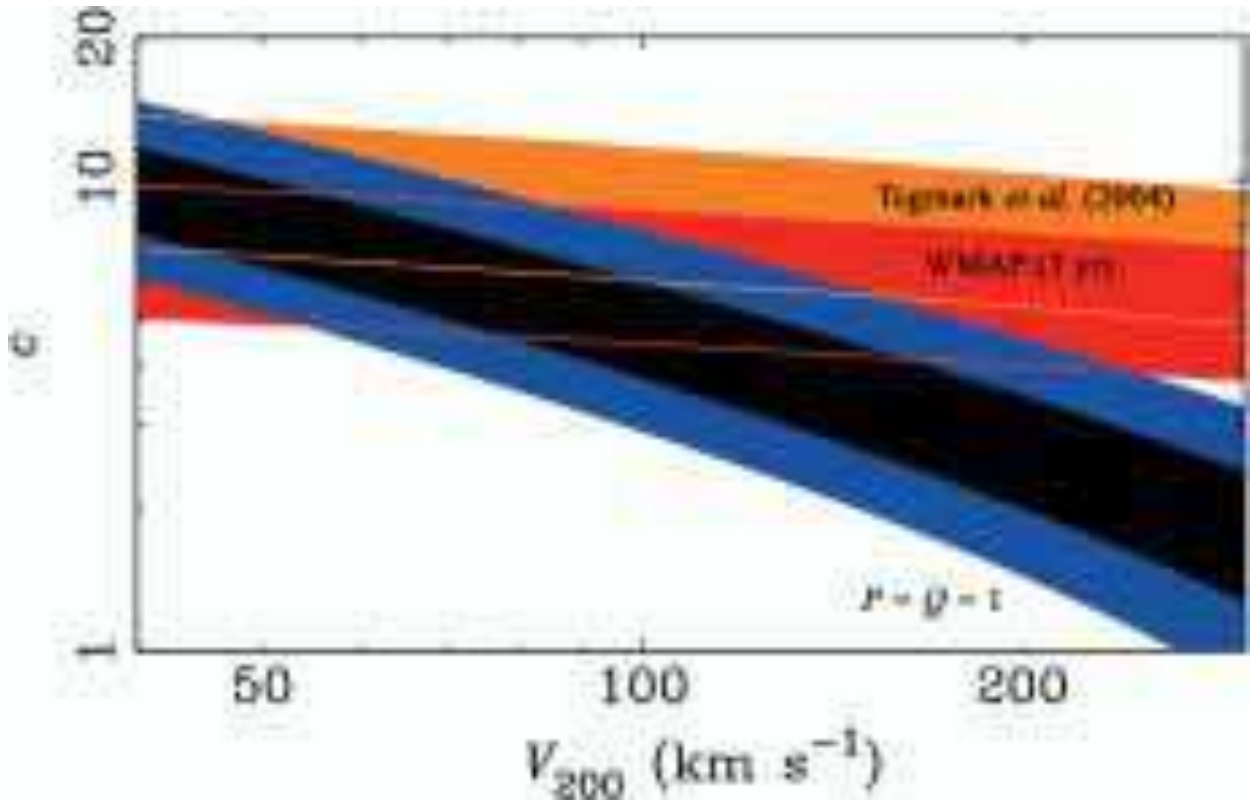}
\caption{The dark matter halo concentration--virial velocity plane.
Blue ($\Pop = 1$) and gray ($\Q = 1$) bands show the region consistent with 
the data for those choices of mass-to-light ratio estimator.  
Orange and red bands show the region that NFW halos are predicted
to occupy for the cosmological parameters of Tegmark \etal\ (2004) and
Spergel \etal\ (2006), respectively.  In both cases, the width of the bands
is $\pm 1 \sigma$.  The predicted bands assume an intrinsic scatter of 
$\sigma_c = 0.14$ (Bullock \etal\ 2001).  The width of the predicted
bands would shrink and the region where they overlap would be reduced 
if we adopted the smaller scatter ($\sigma_c = 0.11$ for relaxed halos) 
advocated by Col{\'i}n \etal\ (2004).  The predicted concentrations 
overlap with the data over a rather narrow range.  This range would 
of course shrink for a smaller intrinsic scatter in $c$.
Theory and data diverge with increasing \VNFW.  
The divergence is most pronounced for the range of halo masses typically 
assumed to be associated with $L^*$ galaxies.
\label{NFWcV}}
\end{figure}

The mismatch between the slopes of the observational and theoretical bands in
Fig.~\ref{NFWcV} is caused by the near universality of the empirical 
dark matter halo found in \S 3.  The data can
be described, albeit crudely, by a single simple form (equation \ref{singlehalo}).
According to NFW, this should not be the case --- halos should spread out
(e.g., top panel of Fig.~\ref{VhQboth}), 
not clump together as observed.  In this case, we
would expect to see more data at high \Vh\ at small $r$.  While there are choices
of \ML\ that have some data in this region, the bulk of the data do not reside
there (Fig.~\ref{DMscale}). 

The slope of the observed bands in Fig.~\ref{NFWcV} is the
result of adjusting the NFW halos concentrations to match the data with 
$\B \equiv 1/2$.  It is worth checking how this changes if we allow the slope to
be free.  A \ML\ choice that leads to a shallower slope and higher amplitude is 
needed to reconcile the data with the model.

Leaving both the slope and the intercept free in the fit, 
we obtain $\C = 1.43$ and $\B = 0.58$ for $\Pop = 1$.  
This steeper slope goes in the wrong direction, as it would make the $\Pop = 1$
band in Fig.~\ref{NFWcV} cut somewhat \textit{more} steeply 
across the predicted band. 
Using the two dozen galaxies for which we have $K$-band data and assuming
a constant $\ML = 0.7\;\Msun/\Lsun$ (the $\Pop = 1$ value for the mean color),
we obtain $\C = 1.53$ and $\B = 0.49$, consistent with $\Q = 1$.
Increasing the stellar mass makes things 
worse, maximum disk being the limiting case 
($\C = 1.17$ and $\B = 0.64$ for $\G = 1$): 
there is too little velocity attributable to dark matter at small radii.  Lowering the
stellar mass within reasonable bounds does not help either.  For $\Pop = 1/2$,
$\C = 1.59$ and $\B = 0.49$.  This a band with precisely the same slope as
illustrated; the change in \C\ only shifts the band a bit to the right.
The case of $\G = 0.4$ (Bottema 1993) gives results indistinguishable
from $\Pop = 1/2$.  To push the stellar mass low enough to get into the right
ballpark, we need $\Pop \lesssim 0.3$ ($\C = 1.65$ and $\B = 0.44$ for 
$\Pop = 0.3$).  This corresponds to an IMF that is a factor of 5 less massive 
than Salpeter.  This is below the practical minimum in stellar mass as determined
from direct integration of the observed IMF (Kroupa 2002).  It is also 
grossly inconsistent with the coherence of the mass discrepancy--acceleration 
relation (McGaugh 2004) and baryonic Tully-Fisher relation (McGaugh 2005), 
the presence of baryonic features observed in rotation curves
(Palunas \& Williams 2000; Sancisi 2004; Kassin \etal\ 2006;
Battaglia \etal\ 2006), observed velocity dispersions 
(Bottema 1997; Kregel, van der Kruit \& Freeman 2005), and
hydrodynamical flows (Weiner \etal\ 2001; Kranz, Slyz \& Rix 2003).
There appears to be no plausible choice of \ML\ that is consistent with 
cosmological expectations and NFW, at least for halos with \VNFW\ 
comparable to the observed circular velocity.  

We have so far implicitly assumed that halo mass scales linearly with 
baryonic mass.  This need not be the case.  One possibility to consider
it that the observed luminosity function of galaxies descends from parent
dark matter halos spanning a smaller range of mass.  This would require a
systematic variation in the fraction of baryons that form the luminous disk
(McGaugh \& de Blok 1998; Okamoto \etal\ 2005), 
but this may already be required to match the observed Tully-Fisher relation
(van den Bosch 2000; Mo \& Mao 2004; see also Navarro \& Steinmetz 2000;
Eke \etal\ 2001; Portinari \& Sommer-Larsen 2006; Dutton \etal\ 2006).

If we drop the assumption that total mass scales with baryonic mass, it is
tempting to conclude that galaxies may be born from halos in the region
allowed in Fig.~\ref{NFWcV}.  This would mean that even massive galaxies
with observed rotation velocities in excess of 200 \kms\ (ranging all the way
up to 300 \kms) would have their origins in relatively modest halos.  
From Fig.~\ref{NFWcV}, we need $\VNFW < 100 \kms$ for vanilla \LCDM, 
and $\VNFW < 180 \kms$ for WMAP third year parameters.
Galaxies would presumably gain their larger observed velocities during the
collapse of their massive disks (see below). 

Examination of Figs.~\ref{VhQboth} and \ref{DMscale} reveals that halos
with $\VNFW \approx 100 \kms$ are about right to explain much of the data.
Unfortunately, such low mass halos do not provide a complete explanation.  
For one thing, there need to be enough baryons available in the halo to form
the observed galaxy.  Assuming a universal baryon fraction of 0.17 (Spergel 
\etal\ 2006), and the most favorable case that all baryons within $R_{200}$
collapse to form the visible galaxy, a halo with $\VNFW = 100 \kms$ has 
enough baryons to make a galaxy with $\Vf = 167 \kms$.  This assumes 
$\Q = 1$ mass-to-light ratios and the baryonic Tully-Fisher relation of
McGaugh (2005).  There is some room
to play with the choice of \ML, but this does little more than assuage the 
extreme assumption that all baryons\footnote{In \LCDM, the
virial radius is closer to $R_{100}$ than $R_{200}$, further reducing the
number of available baryons.} collapse to form the observed galaxy.
It is therefore rather difficult to ascribe the known spiral galaxies with
$200 < \Vf < 300 \kms$ to NFW halos of low mass but normal concentration,
at least for vanilla \LCDM.  This constraint is relaxed for
WMAP 3 year parameters, but even here a halo with $\VNFW = 180 \kms$ 
can give at most enough baryons for a galaxy with $\Vf = 260 \kms$ 
(again taking $\Q = 1$).

Looking in detail,
NFW halos obeying the expected $c$-\VNFW\ relation have more curvature
than observed, predicting too high a velocity at small $r$ and too small a velocity
at large $r$ relative to the data.  Even if we overlook the over-prediction at
$r \sim 5$ kpc, the under-prediction at $r > 20$ kpc is quite serious.  
The compressive formation of a massive disk can raise \Vh, but this process
is much more effective at small radii than at large radii.  In order to explain
the high observed dark halo velocity at large radii, it seems necessary to have
initial halos with at least $\VNFW \approx 200 \kms$.   Such halos grossly
over-predict \Vh\ at small $r$ if they obey the vanilla $c$-\VNFW\ relation.
It may be tempting to simply ignore the data at large radii, as there are relatively 
few with $r > 30$ kpc.  However, there do exist galaxies outside this sample
(e.g., Noordermeer \etal\ 2005; Spekkens \& Giovanelli 2006) with outer
rotation velocities approaching 300 \kms.  These galaxies show no sign of
a turn down in their outer velocities, and obey the same scaling relations as those
in the present sample, so there is no reason to expect them to be any different.  
The need for large \Vh\ at large radii is thus a serious problem.

So far, we have compared the data to the \LCDM\ predictions for primordial
dark matter halos.  It is possible that the process of galaxy formation alters
the halos.  If so, the difference between the predictions and observations might
be interpreted to show what changes occur in this process.  One should bear in
mind that cuspy mass distributions seem to be a generic result of cold collapse
(e.g., Huss \etal\ 1999), and that, once established, it is very hard to unbind the
mass in the deep potential wells of cuspy halos.  Achieving the desired effect is 
thus a tall order from the outset (Sommer-Larsen \& Dolgov 2001; Eke \etal\ 2001).

The primary physical effect which should occur during galaxy formation is the
adiabatic contraction of the dark matter halo in response to the collapse of the
baryonic disk (e.g., Blumenthal \etal\ 1986; Barnes \& White 1984;
Gnedin \etal\ 2004; Sellwood \& McGaugh 2005; Choi \etal\ 2006).
The effect of adiabatic contraction is to further concentrate the
dark matter.  This goes in the wrong direction to help with the problems
discussed here.  However, there are two distinct
aspects of adiabatic contraction to consider.  The slope $\gamma$ will steepen
from its initial value; this goes in the wrong direction.  The velocity
provided by the dark matter also increases.  While this might also seem to act
in the wrong direction, it means that a galaxy of a given rotation velocity
may form in a halo with lower primordial \VNFW.  As discussed above,
this can help, at least in principle.  Unfortunately,
it seems unlikely that the effect can be large enough to help at large radii without 
simultaneously doing far too much at small radii.  Disks heavy enough to have
a substantial impact on outer velocities grossly over-predict inner velocities
if the starting point is an NFW halo (Abadi \etal\ 2003; Sellwood \& McGaugh 2005).
To make matters worse, Dutton \etal\ (2006) argue that a large difference 
between the observed and initial halo velocity scales  
would make reconciling the normalization problem between Tully-Fisher
and the luminosity function practically impossible.  Low mass halos might
also be inconsistent with constraints from weak lensing
(Smith \etal\ 2001; Mandelbaum \etal\ 2006).

After adiabatic contraction, there may be some subsequent
process that further alters the halo.  Perhaps the most common example
is feedback from star formation.  Quite generically, the effects of feedback are 
believed to be stronger in lower mass galaxies owing to their shallower
potential wells (e.g., Thacker \& Couchman 2000).  
Such an effect is unlikely to be sufficiently strong 
(Mac Low \& Ferrara 1999; Gnedin \& Zhao 2002) 
and acts preferentially on the wrong end of the mass function.  
It is the concentrations of high mass galaxies (if not necessarily the 
slope $\gamma$) that are most deviant from the predictions of \LCDM.
If the solution to this problem lies in the processes of galaxy formation,
it likely resides in a rather different direction (see, for example,
El-Zant \etal\ 2001; Weinberg \& Katz 2002; 
Merritt \etal\ 2004; Ma \& Boylan-Kolchin 2004;
Mashchenko, Couchman, \& Wadsley 2006; Tonini, Lapi, \& Salucci 2006;
see also Sellwood 2006).

Another approach would be to alter the initial conditions.  A turndown in
the power spectrum on galaxy scales can reduce halo concentrations.  However,
such a solution would require a rather strange dip in the power spectrum.
Concentrations at small scales are not unreasonable, and also make sense on
larger (cluster) scales (e.g., Vikhlinin \etal\ 2006).  It is the intermediate scale of 
massive spirals, giant ellipticals (Romanowsky \etal\ 2003; Pierce \etal\ 2006), 
and groups (Karachentsev 2005) where the halo concentration 
prediction\footnote{The prediction of a steep inner slope, $\gamma \approx 1$ --- 
the shape rather than the amplitude --- continues to suffer most clearly in 
LSB systems (Simon \etal\ 2005; Kuzio de Naray \etal\ 2006; Goerdt \etal\ 2006;
Sanchez-Salcedo \etal\ 2006).} of \LCDM\ simulations fares worst.

Another possibility is to modify the nature of dark matter.  The cusp-core
problem has already motivated consideration of various possibilities, such as
warm (Bode, Ostriker, \& Turok 2001), fuzzy (Hu, Barkana, \& Gruzinov 2000),
self-interacting (Spergel \& Steinhardt 2000), and meta-cold
(Strigari \etal\ 2006) dark matter, among others 
(e.g., Piazza \& Marinoni 2003; Lee \& Lee 2004; Blanchet 2006).  
It is not immediately clear how any of these proposals fare in the current context.
However, many proposals seem to be meant to produce dark matter halos
with a core.  This in itself is not a complete solution.
One must get the density right, and do so on all relevant scales.  

\section{Conclusions}

We have examined the rotation curves of a large sample of spiral galaxies
in order to constrain the velocity that can be attributed to dark matter.
Taken together, the dark halo velocity can be approximated by
\begin{displaymath}
\log \Vh = \C + \B \log r
\end{displaymath}
with $\B \approx 1/2$ and $\C = 1.47^{+0.15}_{-0.19}$.  
The value of \C\ depends on the choice of stellar mass-to-light ratio, 
with the uncertainty corresponding to the full range of possibilities 
from minimum to maximum disk.  While the dark matter halos of
individual galaxies must differ, this relation gives a tolerable approximation
over the observed range.

In the context of \LCDM, the densities of halos depends on cosmology.
For reasonable cosmological parameters, we expect $\C > 1.6$.
This is only consistent with the data in the limit $\ML \rightarrow 0$,
which is excluded by a wide variety of independent lines of evidence.
The problem is not just that rotation curve data prefer halo models
with a core over those with a cusp.  The amplitude of the velocity 
provided by dark matter is too low at all observed radii.  

We have emphasized here the importance of spiral galaxy dark matter
halo densities being substantially lower than anticipated by \LCDM.  
For halos of $10^{12}\; \Msun$, the density discrepancy is nearly an order 
of magnitude (see also Sellwood 2006).
Getting the halo density in the right ballpark would be an important step 
towards a successful theory.  Ultimately, one must explain in detail
the observed coupling between dark and baryonic components 
(e.g., Sancisi 2004; McGaugh 2004).

\acknowledgements 
We thank the referee for comments which made this manuscript more
clear and led to substantial improvements in the figures.
The work of SSM is supported in part by NASA grant NAG513108.
RKdN and JHK are supported by NSF grant AST0505956.  
This publication makes use of data products from the Two Micron All Sky Survey, which is a joint project of the University of Massachusetts and the Infrared Processing and Analysis Center/California Institute of Technology, funded by the National Aeronautics and Space Administration and the National Science Foundation.

\end{document}